\documentclass{article}

\usepackage{microtype}
\usepackage{graphicx}
\usepackage{booktabs} %
\usepackage{listings}
\usepackage{subcaption}
\usepackage{float}

\usepackage{hyperref}

\usepackage{breakurl}
\usepackage{hyperref}

\newcommand{\tfe}{TensorFlow Eager}
\newcommand{\tf}{TensorFlow}
\newcommand{\tfnamespace}{\texttt{tf}}

\newcommand{\defun}{\texttt{function}}  %

\usepackage[accepted]{sysml2019}

\sysmltitlerunning{TensorFlow Eager:  A multi-stage, Python-embedded DSL for machine learning}

\begin{document}

\twocolumn[
\sysmltitle{\tfe: A multi-stage, Python-embedded DSL for machine learning}

\sysmlsetsymbol{equal}{*}

\begin{sysmlauthorlist}
\sysmlauthor{Akshay Agrawal}{google}
\sysmlauthor{Akshay Naresh Modi}{google}
\sysmlauthor{Alexandre Passos}{google}
\sysmlauthor{Allen Lavoie}{google}
\sysmlauthor{Ashish Agarwal}{google}
\sysmlauthor{Asim Shankar}{google}
\sysmlauthor{Igor Ganichev}{google}
\sysmlauthor{Josh Levenberg}{google}
\sysmlauthor{Mingsheng Hong}{google}
\sysmlauthor{Rajat Monga}{google}
\sysmlauthor{Shanqing Cai}{google}
\end{sysmlauthorlist}

\sysmlaffiliation{google}{Google Brain, Mountain View, CA, USA}

\sysmlcorrespondingauthor{Akshay Agrawal}{akshayka@cs.stanford.edu}
\sysmlcorrespondingauthor{Alexandre Passos}{apassos@google.com}

\sysmlkeywords{Machine Learning, Imperative, Multi-Stage Programming, Dataflow}

\vskip 0.3in

\begin{abstract}
\tfe{} is a multi-stage, Python-embedded domain-specific language
for hardware-accelerated machine learning, suitable for both interactive research and production. 
\tf, which \tfe{} extends, requires
users to represent computations as dataflow graphs;
this permits compiler optimizations
and simplifies deployment but hinders rapid prototyping and run-time dynamism.
\tfe{} eliminates these usability costs
without sacrificing the benefits furnished by graphs: It provides an
imperative front-end to \tf{} that
executes operations immediately and a JIT tracer that translates Python functions
composed of \tf{} operations into executable dataflow graphs.
\tfe{} thus offers a multi-stage programming model
that makes it easy to interpolate between
imperative and staged execution in a single package.
\end{abstract}
]

\printAffiliationsAndNotice{Authors listed in alphabetical order.} 

\section{Introduction}
\label{introduction}
Many contemporary libraries for machine learning share a similar structure: they provide suites of primitive operations and functions to automatically differentiate compositions thereof \citep[see, e.g.,][]{bergstra2010theano, tokui2015chainer, maclaurin2015autograd, chen2015mxnet, abadi2016tensorflow, paszke2017pytorch, 2017gluon, neubig2017dynet, innes18flux, frostig18jax}. These software packages in fact more closely resemble domain-specific languages (DSLs) than libraries \cite{juliaml}. Indeed,  models written using automatic differentiation software are often referred to as \textit{differentiable programs}. 

DSLs for differentiable programming are usually embedded in a host language \citep[for a reference on embedded DSLs, see][]{hudak1996}, and they can be roughly classified as either \textit{imperative} or \textit{declarative}, in the programming languages sense.  Programming in an imperative DSL for differentiable programming is like programming in an imperative programming language such as Python: the construction and execution of primitive operations are inextricably tied, with each operation returning concrete numerical data. While imperative DSLs provide a natural programming paradigm, when embedded in an interpreted language like Python--- which is the case for popular DSLs like Chainer \cite{tokui2015chainer} and PyTorch \cite{paszke2017pytorch}---performance is bottlenecked on the interpreter and serialization of models is difficult. To address these problems, declarative DSLs  separate the specification of models from their execution. These ``define-before-run'' libraries require users to \textit{stage} their models as dataflow graphs, permitting compiler optimizations and the exploitation of parallelism, and simplifying deployment, distribution, and code generation \citep[see, e.g.,][]{ bergstra2010theano, abadi2016tensorflow}. But, because declarative DSLs prevent users from using arbitrary host-language constructs, they have steep learning curves and are not suitable for expressing models with data-dependent structures.

An ideal DSL would offer the flexibility and accessibility of imperative execution along with the many benefits of declarative programming, without either of their costs. It is with this motivation in mind that we present \tfe, a Python-embedded DSL for differentiable programming that lets developers interpolate between imperative and staged computations in a single package. \tfe{} offers a \textit{multi-stage programming} model that lets users rapidly prototype programs and selectively stage parts that they wish to accelerate or serialize. It is implemented as an opt-in extension to \tf, and it can be enabled by calling a single \tf{} library function at program start-up.

To empower machine learning practitioners and researchers to be productive from the start, \tfe{} executes imperatively by default. To reap the benefits of dataflow graphs, \tfe{} provides a Python decorator that traces its Python function in a graph-building context, staging primitive operations to construct a dataflow graph with named inputs and outputs and returning an executable \textit{graph function}. While invoking a graph function is syntactically equivalent to calling the Python function from which it was generated, the execution of graph functions bypasses Python: they are executed using a C++ dataflow runtime or are compiled to generate optimized code for CPUs, GPUs, and ASICs. Graph functions and imperative code share a lexical environment, making it simple to invoke graph functions from imperative code, create graph functions that close over imperatively constructed data, and embed imperative code in graph functions via unstaging annotations.

Our contributions are two-fold:
\begin{itemize}
    \item Our implementation is elegant. \tfe{} can be viewed as a multi-stage front-end to \tf. Imperative and staged \tfe{} code share a single set of primitive operations, kernels, and user-visible APIs. Not only does this sharing result in an easy-to-maintain implementation, it also lets us present a single, coherent API surface to our users that is agnostic to execution mode and lets users enjoy the rich ecosystem of tools developed for \tf.
    \item While we are not the first in the differentiable programming community to recognize the value in bridging imperative and declarative programming, we are among the first to present this line of work in the context of multi-stage programming. This contextualization is a contribution insofar as it clarifies discourse and connects two otherwise separate communities.
\end{itemize}
The remainder of this paper is structured as follows: section \ref{sec:related} surveys related work; \S\ref{sec:design} puts forth our design principles, which prioritize usability and researcher productivity; \S\ref{sec:execution} presents our mutli-stage programming model, with details on automatic differentiation, state, hardware acceleration, distribution, staging, and unstaging; \S\ref{sec:implementation} discusses our implementation; and \S\ref{sec:evaluation} provides a quantitative evaluation of the performance of \tfe{} on machine learning models, demonstrating that imperative \tfe{} can train a ResNet-50 on a single GPU just as quickly as TensorFlow can, staged \tfe{} can train a ResNet-50 on a TPU much faster than imperative \tfe{} can, and that staging yields significant speedups for models with small operations, all with minimal code changes.

\section{Related Work}
\label{sec:related}
In \tfe{}, users must manually stage computations, which might require refactoring code (see \S\ref{sec:multi-stage}). An ideal framework for differentiable programming would automatically stage computations, without programmer intervention. One way to accomplish this is to embed the framework in a compiled procedural language and implement graph extraction and automatic differentiation as compiler rewrites; this is what, e.g., DLVM, Swift for TensorFlow, and Zygote do \cite{wei2017dlvm, swift2018, innes2019zygote}. Python's flexibility makes it difficult for DSLs embedded in it to use such an approach. Some projects, like AutoGraph \cite{wiltschko2018autograph} do operate on Python abstract syntax trees to rewrite imperative code to code that constructs dataflow graphs, but such techniques are out of the scope of this paper.

An alternative to staging computations as graphs for performance is to implement fused kernels. For example, NVIDIA provides fused CuDNN kernels for popular recurrent neural network operations that are dramatically faster than non-fused implementations \cite{chetlur2014cudnn}. This approach, while useful, is difficult to scale, as it requires substantial programmer intervention. 

\tfe{} is not the first Python library to offer a multi-stage programming model. JAX \cite{frostig18jax}, a tracing-JIT compiler that generates code for heterogeneous devices via XLA \cite{xla2017}, provides a similar programming paradigm; MXNet and Gluon also let users interpolate between imperative and staged computations, but at a level of abstraction that is higher than ours \cite{chen2015mxnet, 2017gluon}; and PyTorch is implementing a staging tracer that is similar to ours \cite{pytorchjit}. Outside of differentiable programming, Terra is a Lua-embedded DSL that supports code generation, and the paper in which it was introduced presents a thorough treatment of multi-stage programming that is more formal than ours \cite{devito2013terra}; as another example, OptiML is a Scala-embedded DSL for machine learning with support for staging and code generation but without support for automatic differentiation \cite{sujeeth2011optiml}. Outside of DSLs, there are several projects that provide just-in-time (JIT) compilation for Python, of which Numba \cite{lam2015numba} and PyPy \cite{bolz2009tracing} are two examples. 

Multi-stage programming is a well-studied topic in programming languages; a good reference is \cite{taha2004msp}, and a modern design from which we drew inspiration is Scala's lightweight modular staging \cite{rompf2010lms}. Multi-stage programming is related to staging transformations in compilers and partial evaluation in programming languages, for which \cite{jorring1986} and \cite{jones1993partial} are classic references, respectively.

\section{Design Principles}
\label{sec:design}
Our design strives to satisfy two goals: \tfe{} should be immediately recognizable to Python programmers---for example, users should feel at home exploring APIs and prototyping models in IPython notebooks---and it should also provide a smooth path to testing ideas at scale and deploying models for inference on heterogeneous devices. The first two of the following three  principles are in service of the former goal, while the third is in service of the latter.

\textbf{Privilege imperative execution.}
Because Python is an imperative language, \tfe{} operates in an imperative fashion by default; staged execution is opt-in and often unnecessary (see \S\ref{sec:multi-stage} and \S\ref{sec:evaluation} for details). 

\textbf{Seamlessly embed into Python.}
Whereas writing \tf{} code is an exercise in metaprogramming, imperative execution lets programmers enjoy the full extent of the host language: programmers write Pythonic code, complete with familiar language constructs like native control flow (e.g.,
Python \texttt{if} statements and \texttt{while} loops), recursion, arbitrary data structures, and even \texttt{pdb} breakpoints. And, because
we implement automatic differentiation via tracing (\S\ref{sec:autodiff}), the programmer can differentiate through all these constructs. Host-language integration is more than just syntactic sugar---it greatly simplifies the implementation of data-dependent models like segmental recurrent neural networks and recursive neural networks \cite{kong2015segmental, socher2011parsing}.

\textbf{Stage imperative code as dataflow graphs.}
To leverage the benefits of dataflow graphs, \tfe{} provides a mechanism to trace Python functions and stage their operations as graph functions. The staging workflow is detailed in \S\ref{sec:multi-stage}, and the mechanism is described in \S\ref{sec:staging}. \tf{} graphs come with their own set of design principles, which are presented in \cite{abadi2016tensorflow}.

\section{Execution Model}
\label{sec:execution}
This section presents the pillars of \tfe's execution model. \S\ref{sec:multi-stage} describes imperative and staged execution, presenting a workflow that hybridizes the two; \S\ref{sec:autodiff} describes our trace-based implementation of automatic differentiation; \S\ref{sec:state} specifies how we represent mutable state and how we support serialization; \S\ref{sec:devices} details how \tfe{} supports execution across heterogeneous devices; \S\ref{sec:distribution} presents mechanisms for distributed execution; \S\ref{sec:staging} discusses our tracing JIT in detail; and \S\ref{sec:unstaging} discusses mechanisms for escaping staged computations.

The following terminology will be used in the sequel: a \textit{tensor} is a multi-dimensional, typed array, an \textit{operation} is a primitive, possibly stateful function that takes tensors as inputs and produces tensors as outputs, a \textit{kernel} is a device-specific implementation of an operation, and a \textit{model} is a composition of primitive operations. 

\subsection{Multi-stage programming}
\label{sec:multi-stage}
\tfe{} provides two ways of executing operations:
imperatively or as part of a static dataflow graph. Both execution models have access to the same set of operations and kernels, but they differ in how they dispatch
kernels.

\textit{Imperative execution.} By default, \tfe{} executes operations immediately---library functions such as
\texttt{\tfnamespace.matmul} construct operations 
and then immediately execute their kernels. Under this regime, \tfe{} resembles a NumPy-like library
for hardware-accelerated numerical computation and machine learning.
Calling \texttt{.numpy()} on a tensor fetches a NumPy
array storing the tensor's data, and tensors can be supplied to external libraries like
matplotlib that expect NumPy arrays \citep[for a reference on NumPy, see][]{oliphant2015numpy}. As an example,

\begin{lstlisting}[language=Python,basicstyle=\small\ttfamily,columns=fullflexible]
import tensorflow as tf
tf.enable_eager_execution()

def select(vector):
  A = tf.constant([[1.0, 0.0]])
  return tf.matmul(A, vector)

x = tf.constant([[2.0], [-2.0]])  
print(select(x))
\end{lstlisting}
prints
\begin{lstlisting}[language=Python,basicstyle=\small\ttfamily,columns=fullflexible]
tf.Tensor(
[[ 2.]], shape=(1, 1), dtype=float32).
\end{lstlisting}

\textit{Staged execution.} While imperative execution simplifies prototyping,
the overhead of going back and forth into the Python interpreter limits its performance;
representing computations as dataflow 
graphs before executing them not only removes this bottleneck but also allows for inter-op parallelism and optimizations like constant-folding and buffer reuse. Thus, \tfe{} provides a mechanism to \textit{stage} computations as dataflow graphs. In particular, we provide a decorator, \defun{}, that traces the execution of a Python function, recording all \tf{} operations and the tensors flowing between them in a dataflow graph. \defun{} can be thought of as an opt-in, JIT compiler that generates an optimized polymorphic function for a Python function, creating concrete functions backed by dataflow graphs via a straightforward binding-time analysis at run-time. The analogy to compilers is imperfect because the traces generated by \defun{} only record \tf{} operations and not arbitrary Python code, but it nonetheless provides an approximate mental model. One advantage of this tracing mechanism is that the underlying dataflow graph format does not need to support all the dynamism present in the Python code being traced; as long as the set of operations in the trace does not depend on Python state we can generate a correct trace.

Invoking a callable returned by \defun{} will execute a dataflow graph instead of the corresponding Python function. In fact, graph functions are themselves executed by an operation that takes tensors as inputs and a function name as an attribute, and these operations are automatically constructed and executed for the user. For example, if the \texttt{select} function defined in the previous section were decorated with \texttt{@function}, then \texttt{select(x)} would execute an operation that would in turn execute the appropriate graph function. The dataflow graph runtime, which is written in C++, automatically partitions subgraphs across devices and parallelizes operations when possible. Readers interested in the runtime should consult \cite{abadi2016tensorflow}.

The \defun{} decorator supports code generation via XLA \cite{xla2017}. \tfe{} relies upon XLA to execute code on Tensor Processing Units (TPUs) \cite{tpu2017} (see \S\ref{sec:devices}). In addition to performance and hardware acceleration, dataflow graphs simplify distribution (\S\ref{sec:distribution}) and deployment. Details about the mechanism of \defun{} are provided in \S\ref{sec:staging}.

\textit{A multi-stage workflow.} Many users will find the performance of imperative execution sufficient. Purely imperative \tfe{} can match the performance of graph execution when training models with sufficiently expensive kernels, like ResNet-50 \cite{he2016resnet} (see \S\ref{sec:evaluation}). But when imperative performance disappoints, we recommend the following multi-stage workflow, modeled after \cite{taha2004msp}.
\begin{enumerate}
    \item \textit{Implementation.} Develop, debug, and test a single-stage imperative program.
    \item \textit{Analysis.} Using any profiling tool the user is familiar with, identify performance-critical blocks of operations, and express these blocks as staging-friendly Python functions or callable objects.
    \item \textit{Staging.} Decorate the functions identified in the previous step with \texttt{@}\defun.
\end{enumerate}
With respect to the analysis step, the key fact to keep in mind is that \defun{} is \textit{not} a compiler for arbitrary Python code. Rather, it is a JIT tracer that executes Python functions in a graph-building context and only records operations and tensors. In a graph-building context, operations return symbolic representations of values to be computed instead of concrete values, and non-\tf{} Python code executes normally. Python functions that are amenable to staging are those that, when called in a graph-building context, generate a graph that encapsulates the computation of interest.
This means that if a Python function executes non-\tf{} code, then there might be semantic discrepancies between executing the Python function and executing 
the traced dataflow graph. For example, whereas the Python function
\begin{lstlisting}[language=Python,basicstyle=\small\ttfamily,columns=fullflexible]
def add_noise():
  eye = tf.eye(5)
  randn = np.random.randn(5, 5)
  return eye + randn
\end{lstlisting}
will return a different output every time it is invoked, the dataflow graph generated by \defun\texttt{(add\_noise)} will return the same value every time it is called, since a particular random offset generated by NumPy will be inserted into the graph as a constant. Note that if state is represented in terms of \textit{operations} (e.g., if we  replace the call to \texttt{np.random.randn} with \texttt{\tfnamespace.random\_normal}), we can preserve semantics under this tracing model. As a corollary, if a Python function \texttt{f} has Python side-effects (e.g., every call to it increments a global Python counter), then executing it multiple times will not necessarily be semantically equivalent to repeatedly executing the callable returned by \defun\texttt{(f)}. Python functions must also be resilient to being executed multiple times, as the callable returned by \defun{} might trace its Python function multiple times (see the discussion on polymorphism in \S\ref{sec:staging}).

Because \defun{} generates graphs by tracing and not by source code analysis, it fully unrolls Python \texttt{for} and \texttt{while} loops, potentially creating large graphs. If that is a problem, the programmer might need to replace their loops with the equivalent \tf{} control flow constructs. Similarly, the branches of \texttt{if} statements that are taken during tracing are baked into the emitted graphs. Conditionals that depend on the value of tensors will need to be written using \texttt{\tfnamespace.cond}, and \texttt{while} loops that depend on tensor values will need to be rewritten in terms of \texttt{\tfnamespace.while\_loop}s. Python functions that depend on the values of tensors in complicated ways (e.g., via data structures that depend on the values of tensors) might prove to be prohibitively difficult to stage correctly. In such cases, users might need to refactor their functions into staging-friendly and staging-unfriendly helper functions (see the discussion on escaping staged computations in \S\ref{sec:unstaging} for other options).

Note that staging trades off imperative execution (and therefore interactivity) and Python integration (and therefore run-time dynamism) for performance. It is up to the programmer to decide when this trade-off is acceptable and to use staging annotations judiciously. This trade-off can be diminished by using tools like AutoGraph that operate on abstract syntax trees and rewrite Python control flow to dataflow control flow \cite{wiltschko2018autograph}.

\subsection{Automatic differentiation}
\label{sec:autodiff}

We implement a variant of tracing-based reverse-mode automatic differentiation \cite{baydin2018autodiff}, with a few changes to better support partially staged computation. Our implementation is similar to the implementations of Chainer \cite{tokui2015chainer}, Autograd \cite{maclaurin2015autograd}, and PyTorch \cite{paszke2017pytorch}, but our API allows for more fine-grained control over which computations are traced.

The main user-visible concept in the gradient API is a \textit{tape}. If a tape \textit{watches} a value, operations taking this value as an input will be recorded. It is possible to differentiate any scalar that is computed while a tape is active with respect to any watched value. Tapes are composable data structures: multiple tapes can be active simultaneously, and higher-order gradients can computed by having one tape recording while another tape computes a gradient. Listing \ref{lst:second-derivative} gives an example of nesting tapes to compute a second derivative.

\begin{lstlisting}[language=Python,basicstyle=\small\ttfamily,columns=fullflexible, caption={Tapes can be nested to compute higher-order derivatives.}, captionpos=b, label={lst:second-derivative}]
x = tf.constant(3.0)
with tf.GradientTape() as t1:
  with tf.GradientTape() as t2:
    t1.watch(x)
    t2.watch(x)
    y = x * x
  dy_dx = t2.gradient(y, x)  # 6.0
d2y_dx2 = t1.gradient(dy_dx, x)  # 2.0
\end{lstlisting}

Exposing the tape directly (as opposed to high-level Autograd-like gradient functions) lets users control which parts of the computation are traced for automatic differentiation, which can help limit the run-time overhead incurred in the tracing process.

The tape is tightly integrated with the logic responsible for staging code. The first time a graph function is called when a tape is both active and watching one of its inputs, we build a ``forward'' version of this function that returns any intermediate values needed for the backward step, in addition to its named outputs. As such, there is no meaningful change in the amount of computation or memory needed in the backward pass by staging or unstaging a particular function, leading to more predictable performance. Moreover, this ensures that if a computation was staged in the forward pass, its corresponding backward pass will also be staged. 

Note that gradient computation is itself expressed as a function which executes primitive operations, so it is possible to stage it or not.

\subsection{State}
\label{sec:state}
Like \tf{}, \tfe{} keeps program state in \textit{variables}, restoring a variable's value by assigning to it from a restore operation and periodically saving it to disk by sending its value to a save operation. Variables are useful when implementing models because accessing a variable's value automatically watches it on all active tapes, as shown in Listing \ref{lst:tape-vars}.

\begin{lstlisting}[language=Python,basicstyle=\small\ttfamily,columns=fullflexible, caption={Gradient tapes automatically watch variables; compare this code to Listing~\ref{lst:second-derivative}.}, captionpos=b, label={lst:tape-vars}]
x = tf.Variable(3.0)
with tf.GradientTape() as t1:
  with tf.GradientTape() as t2:
    y = x * x
  dy_dx = t2.gradient(y, x)  # 6.0
d2y_dx2 = t1.gradient(dy_dx, x)  # 2.0
\end{lstlisting}

In \tfe{}, variables correspond to Python objects. Each variable object has its own unique storage that is deleted when Python deletes the object. This is true even for traced computations, where staged \textit{read}, \textit{write}, \textit{save}, and \textit{restore} operations may interact with variables. Staged computations reference variables by unique identifiers, which are no longer usable if the Python variable objects they reference do not exist. This correspondence ensures that \tfe{} state conforms to programmer expectations, stored like any other Python state and accessible through Python identifiers.

\begin{figure}[ht]
\centering
\includegraphics[scale=0.8]{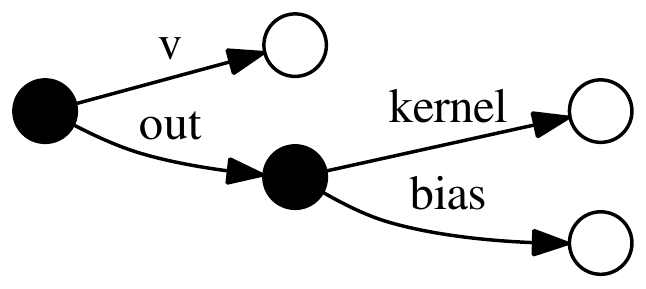}
\caption{Visualization of the dependency graph for Listing~\ref{lst:model-example}, with filled-in intermediate nodes and nodes without fill containing state.}
\label{fig:dependency-graph}
\end{figure}

One challenge when moving from purely staged computation to keeping state in Python objects is matching state between executions of the same program. \tf{} uses unique names for each variable in a program, which relies on the user creating variables in a consistent order. For example creating two copies of the same model requires special consideration when restoring the second model. \tfe{} uses a graph-based matching system, where a directed graph with named edges between objects is serialized along with the program state. On restore, a greedy matching determines a correspondence between serialized Python state and the objects being restored. This matching is local in that it depends only on the objects being saved and restored, not on other parts of the program. Listing~\ref{lst:model-example} and Figure~\ref{fig:dependency-graph} contain a short example.

\begin{lstlisting}[language=Python,basicstyle=\small\ttfamily,columns=fullflexible, caption={Model-building code which implicitly constructs a graph with named directed edges (from attribute names), used for state matching.}, captionpos=b, label={lst:model-example}]
class Net(tf.keras.Model):
  def __init__(self):
    super(Net, self).__init__()
    self.v = tf.Variable(1.)
    self.out = tf.layers.Dense(1)

  def call(self, x):
    return self.out(
      tf.nn.softplus(x * self.v))
\end{lstlisting}

Variables are the most common type of state, but other state is similarly scoped to a Python object and matched as part of a directed graph with named edges. Examples include an iterator over input data whose position in a dataset is serialized, mutable hash tables, and outside of traced code even miscellaneous Python state such as NumPy arrays can use graph-based state matching.

Staging enables serializing the program for use without a Python interpreter, as in \tf{}. A typical development workflow involves using graph-based state matching while writing and tweaking a \tfe{} program, then serializing a trace for use in a production environment that executes the trace using \tf's C++ API.

\subsection{Devices}
\label{sec:devices}

\tfe{} makes it simple to use a variety of devices, such as CPUs, GPUs, and TPUs. During program startup, the runtime detects the devices that are available to the machine, and makes it possible to both execute operations on them and store data on them. Imperative and staged computations use the same underlying \texttt{Device} abstraction, which makes it possible to both execute operations on devices and store data on them. A user-visible API endpoint \texttt{list\_devices} is exposed which lists all devices that the runtime is aware of. 

All tensors exposed to the user are handles to data stored on a particular device. The runtime is also aware of how to copy data between various types of devices, and exposes this functionality through API endpoints on tensor instances.

\begin{lstlisting}[language=Python,basicstyle=\small\ttfamily,columns=fullflexible,caption={Tensor copies between CPU and GPU.},captionpos=b]
a = tf.constant(1.0)  # stored on CPU
b = a.gpu()  # stored on GPU
\end{lstlisting}

When executing an operation, the runtime expects to have a specific device to run the operation on. \tfe{} exposes a context manager, \texttt{device}, so that the user can  control which device operations execute on. The user is not required to use this API, as the runtime is able to select a device based on the availability of kernels. When an operation has inputs on devices different from the device where the operation is executing, the runtime transparently copies the inputs to the correct device. This frees the user from having to explicitly copy tensors between various devices.

\begin{lstlisting}[language=Python,basicstyle=\small\ttfamily,columns=fullflexible, caption={Executing a GPU operation with inputs on the CPU.},
captionpos=b, label={lst:devices}]
# stored on CPU
a = tf.constant(1.0)
b = tf.constant(2.0)

with tf.device("/gpu:0"):
  c = tf.add(a, b) 

assert c.numpy() == 3.0
\end{lstlisting}

Because graph functions are executed via a primitive operation, it is also possible to use the \texttt{device} context manager to run graph functions on various devices. If operations inside the graph function are explicitly placed on another device, they override the outer device context.

Graph functions can serve as a unit of compilation for accelerators; we use this to efficiently execute code on TPUs. When a staged computation is placed on a TPU, \tfe{} automatically invokes XLA to compile the graph and produce a TPU-compatible executable. \tfe{} does make it possible to execute code imperatively on TPUs, but the overhead of compiling operations for TPU and dispatching the generated code is significant. When amortized over a large graph function, this overhead becomes negligible (see \S\ref{sec:evaluation} for a quantitative example). Note that this programming model is similar to JAX \cite{frostig18jax}, which provides a Python decorator that JIT-compiles functions via tracing and XLA. Finally, compiling staged computations through XLA provides us more opportunities for optimization, including layout optimization, instruction scheduling for concurrency, and operation fusion. Techniques like tensor re-materialization can make it possible to fit a staged model into TPU memory when it would be impossible to do so on an operation-by-operation basis.

\subsection{Distribution}
\label{sec:distribution}
The current system supports distributed execution with a single central server running the main (typically Python) program and several worker servers running on remote hosts. Each worker server adds its locally available devices (for example, CPUs, GPUs, or TPUs)  to the pool of devices available to the main program. The main program can then execute operations or whole graph functions on remote devices through the worker servers.

The remote devices are identified by application-level names. The names contain the job name, task inside the job, as well as the specific device available for the task. For example, "/job:training/task:2/device:GPU:0". When a server is brought up to be a part of a cluster, it is given the mapping from the application-level names to specific server instances identified by DNS names or IP addresses.

To run an operation on a remote device, the user uses the same syntax as for local devices (see \ref{sec:devices}) but uses a remote device name instead of the local device name. Tensors produced as the result of running an operation on a remote device stay on the remote device. Users can then either perform more operations on these tensors or copy them to the central server (e.g. to use their value in an \texttt{if} statement).

Some computations running on remote devices can directly communicate and synchronize between each other. In such cases, developers need to start these computations concurrently, e.g. using Python threads.

\subsection{Staging computations}
\label{sec:staging}

The particular type of staging that \tfe{} supports is similar to lightweight modular staging \cite{rompf2010lms}, which in turn is a form of partial evaluation \cite{jones1993partial}. As stated in \S\ref{sec:multi-stage}, we expose a user-visible API endpoint named \defun{} that takes a Python function and returns an object which, when called, executes a dataflow graph created by running the user-provided Python function in a graph-building context. In this section, we discuss the implementation of \defun{} in detail. 

\textit{Polymorphism.}
All Python functions are polymorphic in their inputs. In contrast, graph functions are \textit{not} polymorphic: they have a fixed number of inputs, which are statically typed. We bridge this semantic gap between Python functions and graph functions by implementing a trace cache, similar to the one described in JAX \cite{frostig18jax}. The object \texttt{F = function(f)} maintains a cache mapping from inferred input signatures to concrete graph functions. In particular, each time \texttt{F} is invoked, its inputs are processed and their signature is inferred: tensors are represented as abstract types (numerical type and shape tuples), while non-tensor values are encoded by object identity. This input signature, coupled with a small amount of metadata about the surrounding program state such as the requested device, becomes a key into a cache of graph functions. A cache miss triggers a trace of \texttt{f} on the given inputs, while a cache hit results in the reuse of a previously created graph function. In this sense, \defun{} provides ad hoc polymorphism \cite{strachey2000fundamental} or function overloading.

Not only is specializing functions on input types required for correctness, it also lets us generate optimized graphs --- this kind of optimization is well-known, and, indeed, one of the primary motivations for partial evaluation \cite{jones1993partial, taha2004msp, rompf2010lms}.

Like JAX, \defun{} specializes on the run-time values of non-tensor arguments  to let them parameterize the computation (\defun{} specializes automatically, whereas JAX makes this process a manual one). For example, it is common to write Python functions that take a boolean \texttt{is\_training} argument that determines whether or not dropout is applied. Our implementation of binding-time analysis ensures that graph functions are specialized on the value of the boolean argument (see listing \ref{lst:dropout} for an example).
\begin{lstlisting}[language=Python,basicstyle=\small\ttfamily,columns=fullflexible, label={lst:dropout}, caption={This code transparently makes two graph functions.}, captionpos=b]
@tf.contrib.eager.function
def lossy_matmul(W, x, training=True):
  outputs = tf.matmul(W, x)
  if training:
    outputs = tf.nn.dropout(outputs, 0.2)
  return outputs

W = tf.random_normal((3, 5))
x = tf.random_normal((5, 1))
# Executes a graph with dropout.
lossy_outputs = lossy_matmul(W, x,
  training=True)
# Executes a graph without dropout.
exact_outputs = lossy_matmul(W, x,
  training=False)
\end{lstlisting}
The user also has the option of specifying an input signature to eliminate input polymorphism. In this case, we guarantee that we only generate a single graph function using only the shape and numeric type information specified in the signature. This can be useful for serialization and error-checking, and for creating a single function that can handle arbitrary batch sizes or sequence lengths.

\textit{Lexical closure.}
\defun{} is capable of tracing Python functions that lexically close over tensors or variables --- these closed-over objects are treated as ``captured'' inputs that are silently passed to the graph function at call-time, without programmer intervention. Variables are captured by reference and not by value, which means that graph functions are free to mutate them. Listing \ref{lst:closure} provides an example.

\begin{lstlisting}[language=Python,basicstyle=\small\ttfamily,columns=fullflexible, label={lst:closure}, caption={\defun{} transparently captures closed-over tensors and variables, forwarding them to \tf{} functions as inputs.}, captionpos=b]
v = tf.Variable(0.0)

@tf.contrib.eager.function
def mutate():
  v.assign_add(1.0)
  return v.read_value()

mutate()
assert float(v.read_value()) == 1.0
v.assign_add(1.0)
assert float(v.read_value()) == 2.0
mutate()
assert float(v.read_value()) == 3.0
\end{lstlisting}

\textit{Composition.}
Because graph function execution is implemented as an operation, graph functions compose naturally: the graph of a function may include a function-call operation that executes another function. For example, consider the following code block:

\begin{lstlisting}[language=Python,basicstyle=\small\ttfamily,columns=fullflexible, label={lst:nestedfns}, caption={Graph functions can be nested.}, captionpos=b]
@tf.contrib.eager.function
def inner(a):
  return tf.nn.relu(a)
  
@tf.contrib.eager.function
def outer(a, b):
  return inner(tf.matmul(a, b))

outer(tf.eye(3), tf.diag([-1.0, 1.0, 2.0]))
\end{lstlisting}

\begin{figure}
\centering
\begin{subfigure}[t]{.2\textwidth}
    \includegraphics[scale=0.5]{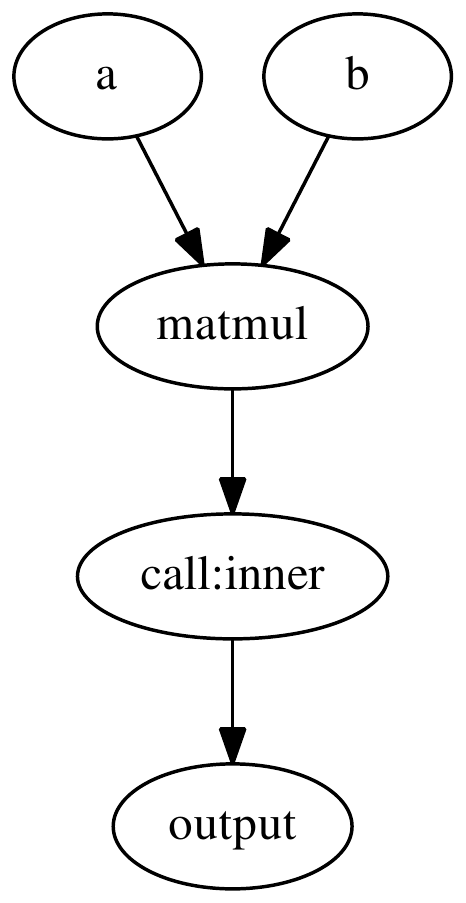}
    \caption{The graph generated for \texttt{outer}; note the \texttt{call} operation that executes \texttt{inner}'s graph function.}
\end{subfigure}\hfill%
\begin{subfigure}[t]{.2\textwidth}
    \includegraphics[scale=0.5]{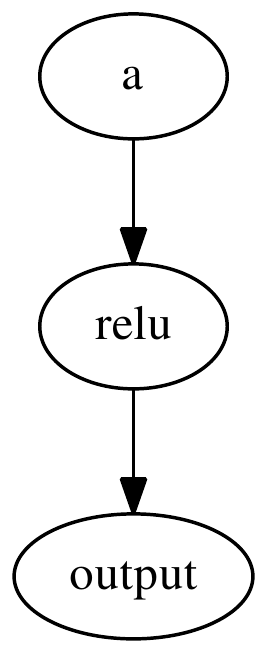}
    \caption{The graph generated for \texttt{inner}.}
\end{subfigure}
\caption{\defun{} composes; above, the graphs for Listing \ref{lst:nestedfns}.}
\label{fig:nestedfns}
\end{figure}

The call to \texttt{outer} will generate two graph functions, one for \texttt{inner}, and one for \texttt{outer} that contains a call to \texttt{inner}'s graph function. Figure \ref{fig:nestedfns} shows what their corresponding graphs look like. 

\textit{State creation.}
When building machine learning models, it is common to write Python functions that create and initialize variables the first time they are called. To support this idiom, \defun{} imposes some requirements on the decorated function \texttt{f}. State, such as TensorFlow variables, must only be created the first time \texttt{f} is called; how that is accomplished is left to the implementation of \texttt{f}. If any variables are created in the first execution of \texttt{f}, then \defun{} will trace \texttt{f} a second time to record the behavior that will be used from then on. No variables may be created during that second trace, or any subsequent one.

\subsection{Escaping staged computations}
\label{sec:unstaging}

\textit{Embedding imperative code in graphs.} As discussed in \S\ref{sec:multi-stage}, staging computations requires the programmer to refactor the to-be-staged code into Python functions that, when traced, construct dataflow graphs. This process may at times seem prohibitively difficult, as it can require replacing complicated Python control flow with \tf{} control flow or even implementing custom operations along with custom C++ kernels --- indeed, this observation was one of the motivations for building \tfe{} to begin with.

For concreteness, say that we have a Python function that we wish to stage, and say that the function is almost entirely staging-friendly with the exception of a call to a data-dependent recursive Python function that performs some operations on tensors. In this case, we have three options: we can refactor the function into three functions, staging the code before and after the recursive call and leaving the recursive call unstaged; we can give up on staging the function if refactoring proves too onerous; or we can stage the entire function while wrapping the recursive call in a \texttt{py\_func}, an operation that takes a Python function as an attribute and executes it imperatively, even in the context of staged code. 

\texttt{py\_func} executes its Python function under a gradient tape (see \S\ref{sec:autodiff}) and as such it is differentiable; it also has both CPU and GPU kernels. When executing in imperative mode, wrapping a Python function in a \texttt{py\_func} has essentially no effect. But, in staged computations, i.e. in dataflow graphs, the \texttt{py\_func} operation is a way to embed imperative, Pythonic code into a dataflow graph. Equivalently, \texttt{py\_func} can be viewed as a way to quickly implement  custom operations using Python instead of C++.

The benefit of \texttt{py\_func} is that it makes it easier to decorate large Python functions with \texttt{@function}. Disadvantages include a potential performance hit, as \texttt{py\_func} returns control to a single-threaded Python interpreter, and the fact that graphs with \texttt{py\_func}s are not in general serializable.

\textit{Escaping traces.}
We provide a Python context manager, \texttt{\tfnamespace.init\_scope}, that pauses the trace and jumps into the imperative context. We use this scope to implement \defun's state-creation contract; most users, on the other hand, will have no use for it.
    
\section{Implementation}
\label{sec:implementation}
We have implemented the design presented in \S\ref{sec:execution}, and all of our code is open source\footnote{\url{https://github.com/tensorflow/tensorflow}}. Because \tfe{} was built as an extension to \tf, the implementation is not large: staging is implemented in approximately 2000 lines of Python, automatic differentiation is split across 900 lines of Python and 600 lines of C, and the imperative runtime---i.e., the code responsible for constructing and executing operations---is implemented in approximately 4000 lines of C++. \tfe{} also provides a lightweight C API that exposes our runtime, and several of our colleagues are using this API directly in their own projects. 

\tfe{} inherits the benefits of \tf's implementation. In particular, \tfe{} is cross-platform, running on the Linux, Mac OS X, Windows, Android, and iOS operating systems, and various x86, ARM, and NVIDIA GPU architectures; it executes staged computations using a dataflow executor that can run over ten thousand subgraphs in parallel and that runs kernels in parallel when possible, across multiple CPU cores or GPU streams; it provides high-level Python APIs for training models and C++ APIs for inference \citep[see][\S5]{abadi2016tensorflow}. \tfe{} also provides access to the over 900 primitive operations that \tf{} offers.

\tfe{} and \tf{} differ slightly but significantly in their implementations of staged execution. In \tf, the dataflow graph defines the union of all the computations that the author of the graph might be interested in; the actual computation to execute is defined when the programmer requests the runtime to fetch the concrete values of some set of tensors resident in the graph. This amounts to a discrepancy between what is expressed in Python and what is executed by the \tf{} runtime. To provide a more Pythonic programming model, \tfe{} represents each staged computation as a graph function, i.e., a graph with named inputs and outputs, representing the \textit{exact} computation of interest. This approach still allows for graph optimizations: for example, non-stateful operations that are not reachable from the outputs of a function are pruned, just as in \tf.

Graph functions provide benefits outside the realm of usability as well. Because graph functions are executed via an operation, we get function composition for free. In the context of single-coordinator distributed training, in which a single subgraph is executed by $N$ workers, graph functions can reduce memory pressure on the coordinator: the coordinator only needs to own a graph function that contains $N$ function-call operations (instead of $N$ copies of a subgraph).

\section{Evaluation}
\label{sec:evaluation}

\tfe{} considerably simplifies rapid prototyping. This at times trades off execution speed for development ease. In this section, we present examples\footnote{The code for these example models and others is available at \url{https://github.com/tensorflow/tensorflow/tree/master/tensorflow/contrib/eager/python/examples}.} showing how we can use \defun{} to recover the speed of \tf{}.

\textit{Experimental setup.} The benchmarks were run within a docker container on a machine with an Intel(R) Xeon(R) W-2135 CPU with 12 cores at 3.7GHz, 64GB of memory, and a GTX 1080 GPU with 8GB of memory. The TPU benchmark was run on a publicly available Cloud TPU. Each benchmark run was 10 iterations, and an average of 3 runs was reported. For staged computations, build and optimization times were not included as these are one-time costs that are usually amortized over a number of runs.

\begin{figure}
    \centering
    \includegraphics[width=8cm, height=5cm]{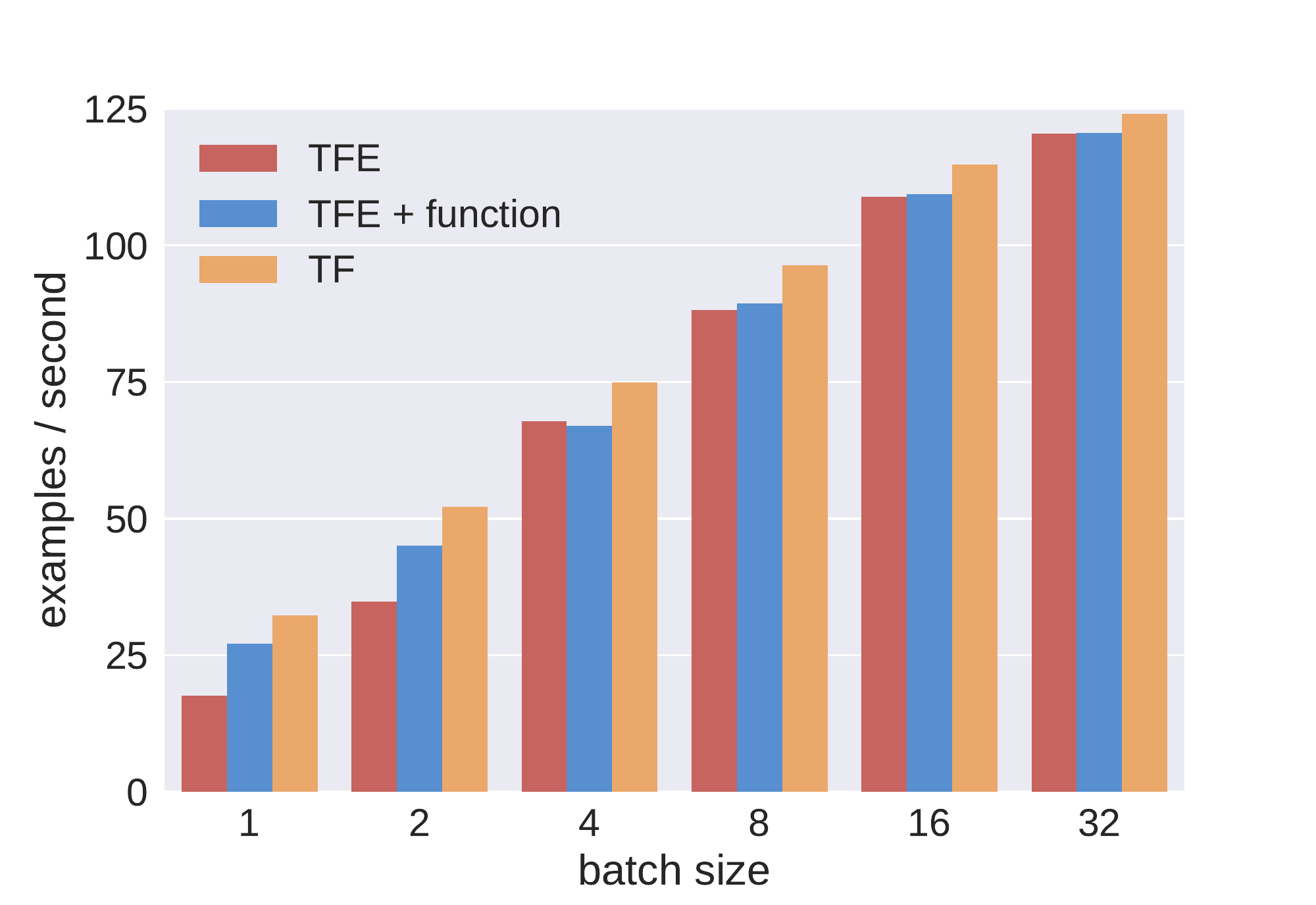}
    \includegraphics[width=8cm, height=5cm]{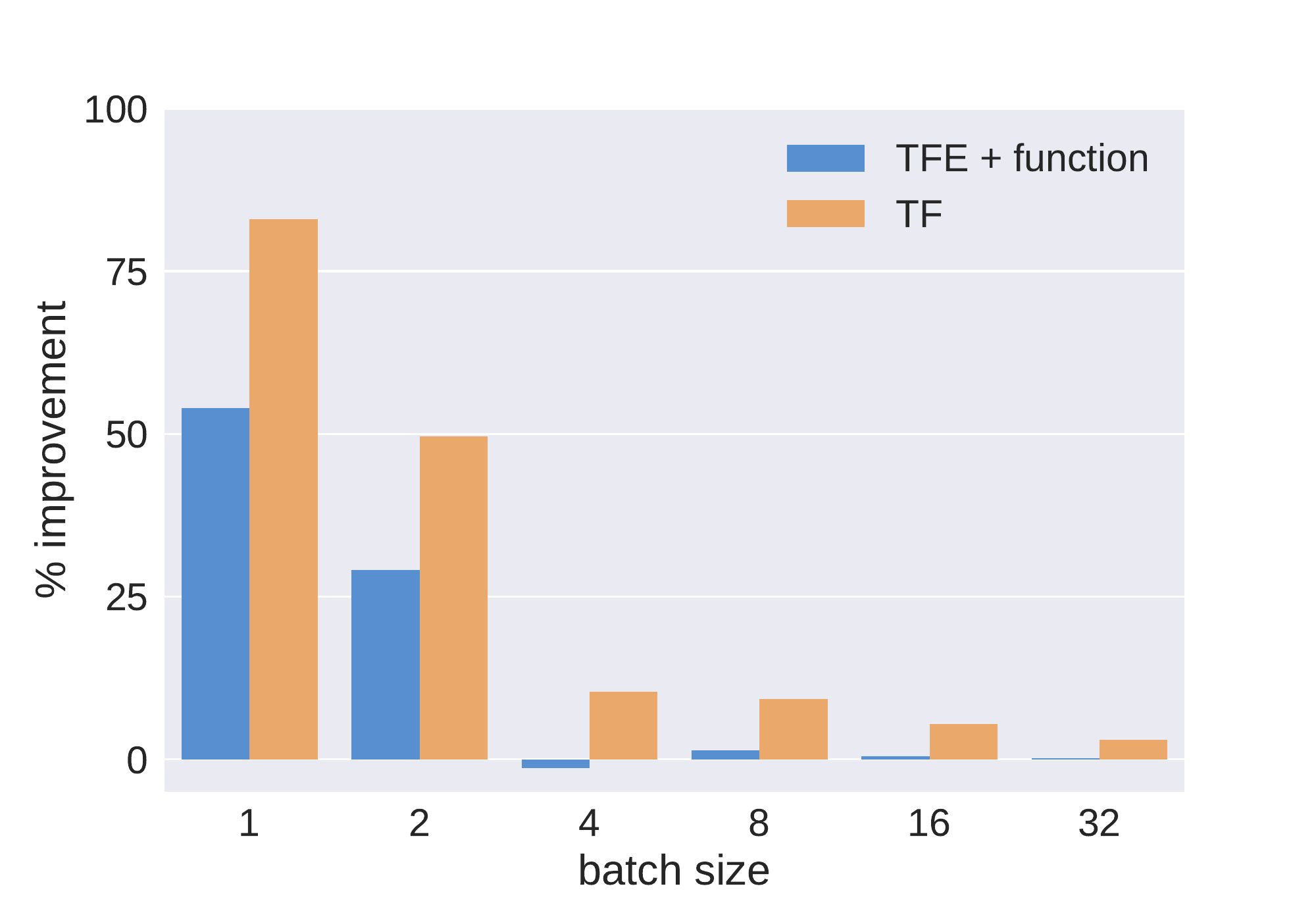}
    \caption{Examples per second when training ResNet-50 on a GPU (top). Percent improvement over \tfe{} (bottom).}
    \label{fig:gpu_benchmark}
\end{figure}

\textit{ResNet-50.} In Figure \ref{fig:gpu_benchmark} we show the performance of training a ResNet-50 model, comparing \tfe{}, \tfe{} with the forward pass and gradient application staged with \defun{}, and \tf{}. The top chart shows the raw examples per second, and the bottom chart shows the improvement that \tfe{} with \defun{} and \tf{} show over \tfe{}. For smaller batch sizes, staging computations yields significant speed-ups. These speed-ups vanish as the batch size increases, since the ratio of the time spent in kernels over the time spent in Python increases. Additionally, training a ResNet doesn't benefit significantly from inter-op parallelization, so the staged computation is effectively as serial as the unstaged computation. These performance characteristics should hold true for other sufficiently large models, i.e., imperative performance will often be similar to staged performance. The code used to generate these benchmarks all rely on the same \texttt{Model} class; converting the code to use \defun{} is simply a matter of decorating two functions.

\textit{ResNet-50 on TPU.} It is possible to run single operations on a TPU using \tfe{}. The performance of training ResNet-50 on ImageNet \cite{imagenet_cvpr09} using \tfe{} versus \tfe{} with \defun{} is shown in Table~\ref{tab:tpu_resnet_benchmark}. Training the model in a per-operation fashion is slow, even at a batch size of 32; staging yields an order of magnitude improvement in examples per second.

\begin{table*}
    \centering
    \begin{center}
    \begin{tabular}{ l c c c c c c }
    & 1 & 2 & 4 & 8 & 16 & 32 \\ 
    \hline
    \tfe{} & 1.06 & 1.99 & 4.3 & 8.4 & 16.6 & 30.3 \\  
    \tfe{} with \defun{} & 21.7 & 42.6 & 83.9 & 165.8 & 197.7 & 241.9 \\  
    \end{tabular}
    \end{center}
    \caption{Examples per second training ResNet-50 on a TPU.}
    \label{tab:tpu_resnet_benchmark}
\end{table*}

An important caveat is that these benchmarks do not exploit the hardware optimally. They are presented as illustrative of how staging lets us target accelerators like TPUs with practically no code changes. We don't present an accompanying \tf{} benchmark for this reason.

\begin{figure}[]
    \includegraphics[width=8cm, height=5cm]{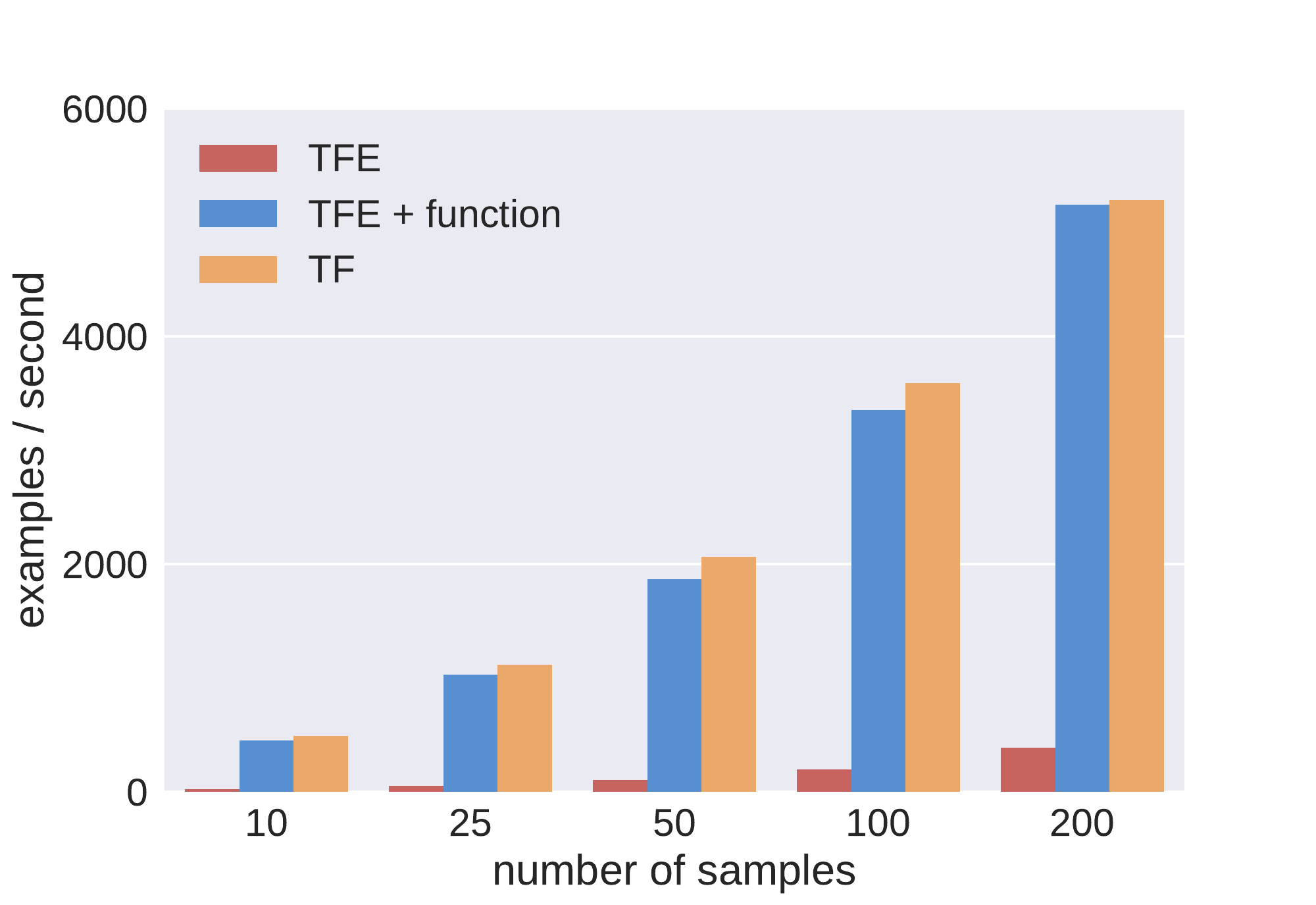}
    \caption{Examples per second training L2HMC on a CPU.}
    \label{fig:l2hmc_cpu}
\end{figure}

\textit{L2HMC.} In Figure \ref{fig:l2hmc_cpu} we show performance of an L2HMC \cite{levy2017l2hmc} implementation, comparing \tfe{}, \tfe{} with \defun{}, and \tf{} on synthetic data running on the CPU. The benchmark samples from a 2-dimensional distribution, with 10 steps for the leapfrog integrator. This example highlights the trade-off between debuggability and performance: by bypassing Python overheads and via buffer reuse and other static optimization, staging increasing examples per second by at least an order of magnitude. And while the trade-off exists, it is not particularly onerous here --- simply decorating a single function recovers the full performance of \tf{}. This benchmark stages computation aggressively, essentially running the entire update as a graph function. Depending on the desired visibility into the model's execution during development, it is possible to stage less aggressively.

\textit{Note.} These examples were chosen as they lie at opposite ends of the tradeoff between execution speed and development speed. We expect most real-world models to fall somewhere between these two, and to be able to recover performance by staging as required. \tfe{} is an evolving technology, and closing the gap between imperative and staged performance is being worked on.

\section{Conclusion}
\label{sec:conclusion}

We presented \tfe{}, an extension to \tf{} that makes what was once a declarative DSL for differentiable programming into a multi-stage, imperative-first one. \tfe{}'s imperative-by-default behavior makes it suitable for beginners and researchers alike, and the option to stage computations as graph functions lets users trade off  the interactivity and Python integration furnished by imperative execution for the benefits provided by static graphs, performance and ease of serialization among them.

Within Alphabet, dozens have adopted \tfe{}. For example, some researchers use it to implement dynamic language models and reinforcement learning methods, and several internal workshops on \tfe{} have been attended widely. Multiple groups are restructuring their machine learning frameworks to make \tfe{} the default way of using them (examples include libraries for probabilistic machine learning and reinforcement learning), and at least one large research group has engineers dedicated to supporting \tfe. Externally, some university courses have included \tfe{} as part of their curriculum, and 48 percent of respondents to a survey distributed at the 2018 TensorFlow Developer Summit agreed with the statement, ``[\tfe{}]{} is important to me as an iterative development and debugging tool.'' 

\tfe{} is an evolving technology. While it is well-suited for research and pedagogy alike, we are still working to provide an out-of-the-box solution for imperatively-driven distributed training. And while multi-stage programming is powerful ---  wrapping large Python functions in \defun{} often ``does the right thing'' --- staging computations with dynamic control flow can require nontrivial programmer intervention. We hope to decrease this friction via Autograph \citep{wiltschko2018autograph}.

Finally, \tfe{} has informed the evolution of \tf{} itself: 
the upcoming TensorFlow 2.0 uses our implementation to provide an imperative-first, multi-stage programming model similar to the one outlined in this paper.

\section*{Acknowledgements}
We'd like to thank everyone on the TensorFlow team for their feedback and help with the design and implementation of this system. Alex Wiltschko, Pierre Sermanet, Xin Pan, Yaroslav Bulatov, Manjunath Kudlur, and Yuan Yu contributed significantly to motivating and early prototyping of TF Eager. Early users like Sergio Guadarrama, Daniel Abolafia, David Berthelot, Chen Li, Debidatta Dwibedi, among others, were key in helping us shape the requirements of this system. A lot of it wouldn't look like it does now without important feedback from DeepMind, especially from Aedan Pope and Tom Hennigan. Fran\c{c}ois Chollet was very helpful in integrating TF Eager with Keras.

\clearpage
\bibliography{tfe}
\bibliographystyle{sysml2019}

\end{document}